\documentstyle[prl,aps,multicol,epsf]{revtex}
\begin{document}
\title{
Quantized acoustoelectric current in a finite-length ballistic quantum
channel:\\ The noise spectrum }

\author{ Y. M. Galperin$^{(a)}$, O. Entin-Wohlman$^{(b)}$ and
 Y. Levinson$^{(c)}$ }

\address { $^{(a)}$ Department of Physics, University of Oslo, Box
1048 Blindern, N-0316 Oslo, Norway \\ and Solid State Division,
A. F. Ioffe Physico-Technical Institute, 194021 St. Petersburg,
Russia\\ $^{(b)}$ School of Physics and Astronomy, Raymond and Beverly
Sackler Faculty of Exact Sciences, \\Tel-Aviv University, Tel-Aviv
69978, Israel \\ $^{(c)}$Department of Condensed Matter Physics, The
Weizmann Institute of Science, Rehovot 76100, Israel}

\date{\today}

\maketitle
\begin {abstract}
Fluctuations in the acoustoelectric current, induced by a surface
acoustic wave propagating along a ballistic quantum channel, are
considered. We focus on the large wave-amplitude case, in which it
has been experimentally found that the current is quantized, and
analyze the noise spectrum. A phenomenological description of the
process, in terms of a random pulse sequence, is proposed. The
important ingredients of this description are the probabilities,
$p_{n}$, for a surface acoustic wave well to capture  $n$
electrons. It is found that from the noise characteristics one can
obtain these probabilities, and also estimate
 the typical length scales of the regions in which the electrons
are trapped.

\end {abstract}
\begin{multicols}{2}

Acoustic methods appear to be extremely useful in probing physical
properties of the two-dimensional electron gas (2DEG) \cite{Wix1}. Surface
acoustic waves (SAW's) induce long-range electric fields, with the spatial
and temporal periodicities of the wave. These fields, which penetrate the
2DEG without any galvanic leads, facilitate a probeless diagnostics. In
addition, the energy loss in such experiments takes place inside a very
small volume which is well thermally coupled to the substrate. As a
result, SAW's usually do not cause overheating of the electronic device.

Two effects are customarily being studied. The first is the {\em
attenuation} of the SAW's due to electric currents induced in the 2DEG.
This effect, which is {\em linear} in the acoustic amplitude,  allows the
determination of the linear response of the 2DEG to an ac perturbation, of
frequency and wave vector of the acoustic wave. For example, studies of
the acoustic attenuation and the velocity in the presence of an external
magnetic field perpendicular to the 2DEG have been exploited to
investigate the ground state of two-dimensional  quantum Hall effect
systems \cite{Wil}. The second is the {\em acoustoelectric effect}, which
a {\em nonlinear} response. In a quantum channel, this effect has been
measured in Ref. \cite{Pepper}. The electric fields induced by SAW's at
the 2DEG drag the electrons, and consequently produce a dc current in a
closed circuit, or a dc voltage across an open circuit.

Recently it has been experimentally demonstrated
\cite{Sh1,Sh2,Ta97} that under proper conditions the
acoustoelectric current through a non-biased pinched-off channel
in a 2DEG consists of a set of plateaus, when the gate voltage, or
alternatively the SAW amplitude, is varied. Below a certain
threshold of the SAW intensity the current is very small (the
structure of the current in that regime has been discussed in
Refs. \cite{Sh1,AE1,AE2}); at the threshold it ``jumps'' to a
quantized value, $ef$,  (where $e$ is the electron charge and $f$
is the SAW frequency), which it keeps up to a second threshold;
then it ``jumps'' again to a second quantized value $2ef$.

Quantization of the current carried by  electrons trapped in a
moving potential has been first addressed in Ref.~\cite{Thouless}
where it has been shown that the current induced by a slowly
moving periodic potential can be quantized in units of $e L/v$
where $L$ is the period of the potential profile and $v$ is its
velocity. Previous theoretical considerations have addressed  the
microscopic origins of the quantized current
\cite{Aizin,Maksym,Flensberg}. Here we also consider the
relatively large SAW amplitude case, but propose a
phenomenological description of the quantized current and its
noise, in terms of a random pulse sequence. The picture we have in
mind stems from the quantitative explanation outlined in Refs.
\cite{Sh1,Sh2,Ta97}.

In the absence of the SAW,  the quantum channel in the 2DEG is
pinched-off, that is, the common Fermi level $\epsilon_{F}$ in the
(non-biased) source and drain is below the bottom of the channel
band, $\epsilon_{B}$, and so electrons from the terminals cannot
penetrate the channel  (see Fig.~1a). When a SAW is propagating
along the channel, the bottom  of the channel band is modulated by
the moving piezoelectric potential profile, which consists of
wells separated by barriers. Such a modulation does not take place
in the terminals because the 2DEG strongly screens the effect of
the wave.  The potential profiles at different times are
schematically depicted in Fig.~1b,c. For a strong enough SAW, the
modulation amplitude $V_{\text{SAW}}$ exceeds the difference
$\epsilon_B -\epsilon_F$. Then the bottoms of the wells that
appear near the source during half of the period of the SAW, are
located below the Fermi level (see Fig.~1b) . Such
 a well may trap one or
more electrons from the source into  bound states $\epsilon_0$ in the well
and then drag them along toward the drain. Upon the arrival of the well at
the drain, the electron energies $\epsilon_0$ are below $\epsilon_F$ (see
Fig.~1b). However, the electrons are adiabatically excited to
$\epsilon_{F}$  because of the squeezing  of the well as it disappears
into the drain, (compare the levels in the right well in Fig.~1b and
Fig.~1c) and the electrons are eventually absorbed by the drain.

Because of the strong Coulomb repulsion between two electrons on the same
localized state, it is plausible that a not-too-deep well will capture
only a single electron, a deeper well will accommodate two electrons, and
so on. Hence, ignoring for the time-being any randomness or fluctuations
in the process, the current induced by a moderate-intensity SAW can be
viewed as a periodic sequence of non-overlapping pulses, whose period is the same as that
of the SAW, $f^{-1}$, with each pulse, $i_{1}(t)$, carrying a single
electron. On the average, such a sequence carries the current $ef$; as the
intensity of the SAW's is increased and their wells become deeper, then
pulses  capable of carrying two electrons, $i_{2}(t)$,  appear, leading to
an average current 2$ef$, and so on. In this way, one obtains a
qualitative picture of the phenomenon of the acoustoelectric current
quantization \cite{Sh1,Sh2,Ta97}.
\begin{figure}[h]
\epsfxsize=6truecm
\centerline{
\epsffile{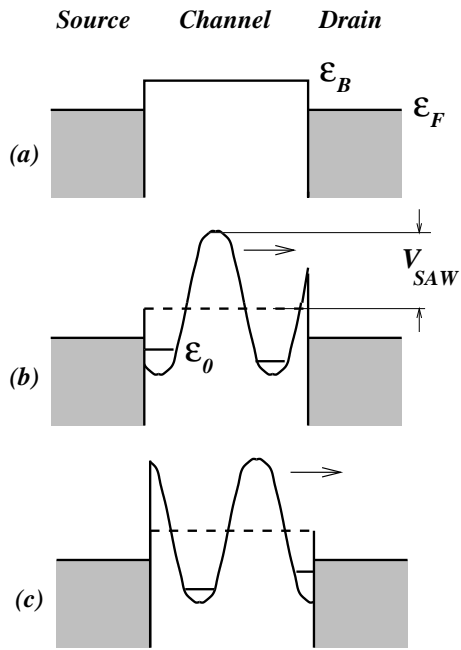}
}
\vspace*{1mm}
 \caption{ Schematic profiles of the potential created
by the SAW at different times.   \label{fig1}}
\end{figure}

 However, naturally the acoustoelectric current is subject to fluctuations.
Fluctuations  will occur since some of the potential wells will not behave
according to the scheme outlined above. For example, near and above the
first threshold, there may be empty potential wells, which do not carry
electrons; near the second threshold there will be wells carrying only one
electron, and the current will be a random sequence of the pulses
$i_{1}(t)$ and $i_{2}(t)$.

We hence propose a description in which the acoustoelectric current
consists of a random sequence of pulses. Let $p_{n}$ be the probability
that a potential well near the source captures $n$ electrons, and
therefore contributes to the current the pulse $i_{n}(t)$, such that
$\int_{-\infty}^{\infty} dt \, i_{n}(t)=ne$. We will show that the
probabilities $p_{n}$, which depend on the temperature and on the
intensity of the SAW's, can be extracted from the {\em low-frequency
fluctuations} of the acoustoelectric current. These probabilities can be
therefore directly probed by the experiment.  Moreover, we will find that
the fluctuation spectrum also provides  information regarding the {\em
characteristic size}, $a_n$,  of the $n$-th electronic state confined in
the well.

In our model, the current is thus a stationary random function of time,
\begin{eqnarray}
{\mathrm I}(t)=\sum_{\nu}{\mathrm F}_{\nu}(t-t_{\nu}),
\end{eqnarray}
with F$(t)$ being any of the pulses $i_{n}(t)$, with probability $p_{n}$,
$\sum_{n}p_{n}=1$, and $t_{\nu}$ denotes the arrival time of the $\nu$th
pulse at the drain. The average time interval between the pulses, $\langle
\tau\rangle$, is determined by the SAW period, i.e., $\langle\tau\rangle
=f^{-1}$.

 We next calculate the average current, and its fluctuation
spectrum. To this end, it will be convenient to  define the Fourier
transform of a single pulse,
\begin{eqnarray}
{\mathrm F}_{\nu}(\omega )=\int_{-\infty}^{\infty} \frac{dt}{2\pi}{\mathrm
F}_{\nu}(t)e^{i\omega t}. \label{fourier}
\end{eqnarray}
Consider first the average current. One has
\begin{eqnarray}
\langle{\mathrm
I}(t)\rangle&=&\left\langle\int_{-\infty}^{\infty}d\omega
e^{-i\omega t} {\mathrm F}(\omega )\right \rangle \, \left \langle
\sum_{\nu}e^{i\omega t_{\nu}}\right \rangle\, ,
\end{eqnarray}
as the pulse shapes and the arrival times are assumed to be statistically
independent. The second average here is carried out over a time interval
$T$ much longer than the duration of the pulses \cite{Rytov}, yielding
$\langle \exp (i\omega t_{\nu})\rangle = 2\pi\delta (\omega )/T$. The
first average then gives
\begin{eqnarray}
 N\sum_{n}p_{n}\int_{-\infty}^{\infty}dt\;i_{n}(t)=Ne\sum_{n}p_{n}n,
\end{eqnarray}
 where $N$ is the total number of pulses, such
that $\langle\tau\rangle =T/N$. As a result, the average of the random
stationary current is
\begin{eqnarray}
\langle{\mathrm I}(t)\rangle =\frac{e}{\langle\tau\rangle}\sum_{n}p_{n}n.
\end{eqnarray}
As is mentioned above, the probabilities $p_{n}$ are functions of
the SAW intensity. We hence conclude that below the first
threshold of the measured current, the dominant probability is
$p_{0}$; at the threshold $p_{1}$ becomes comparable to $p_{0}$,
and above it, at the plateau, $p_{1}$ increases on the expense of
$p_{0}$, and so on.

The noise spectrum of the current, ${\mathrm S}(\omega )$, is 
defined as (see  e. g. Ref.~\onlinecite{Kogan}),
\begin{eqnarray}
{\mathrm S}(\omega )=2\int_{-\infty}^{\infty}
dt \, e^{i\omega t}
\langle \delta{\mathrm I}(t)\delta {\mathrm I}(0)\rangle ,
\end{eqnarray}
with $\delta{\mathrm I}(t)={\mathrm I}(t)-\langle{\mathrm I}\rangle $. To calculate
this quantity we follow Ref. \cite{Rytov}.
 One then finds
\cite{end1}
\begin{eqnarray}
{\mathrm S}(\omega )&=&\frac{8\pi^2 N}{T}\langle |{\mathrm
F}(\omega )|^{2}\rangle \nonumber\\ &+& \frac{8\pi^2}{T}|\langle
{\mathrm F}(\omega )\rangle |^{2} \left \langle \sum_{\nu'\neq
\nu}e^{i\omega (t_{\nu}-t_{\nu '})}\right \rangle \,  .
\end{eqnarray}
The last average here requires the distribution function of the arrival
time differences (which has the average $\langle\tau\rangle $). Denoting
it by $w(\tau )$, and introducing
\begin{eqnarray}
\phi (\omega )\equiv \left \langle e^{-i\omega \tau}\right \rangle
=\int^{\infty}_{0}d\tau\,  e^{-i\omega\tau}w(\tau ), \label{phi}
\end{eqnarray}
it is straightforward to find
\begin{eqnarray}
\left \langle \sum_{\nu' \neq \nu}e^{i\omega (t_{\nu}-t_{\nu
'})}\right \rangle =2N\, \mathrm {Re}\, \frac{\phi (\omega
)}{1-\phi (\omega )},
\end{eqnarray}
in the $N\rightarrow\infty $ limit.

The result for the current fluctuation spectrum is conveniently presented
in the form
\begin{eqnarray}
{\mathrm S}(\omega)={\mathrm S}_{b}(\omega)+{\mathrm S}_{c}(\omega)\,. \label{s}
\end{eqnarray}
Here, ${\mathrm S}_{b}(\omega )$ is a smooth ``background" noise, which is not
affected by the distribution of intervals between pulses
\begin{eqnarray}
{\mathrm S}_{b}(\omega )=\frac{8\pi^2}{\langle\tau\rangle} \Bigl ({\mathrm K}(\omega
)-|{\mathrm H}(\omega )|^{2}\Bigr ),\nonumber\\ {\mathrm K}(\omega )=\langle |{\mathrm
F}(\omega )|^{2}\rangle ,\qquad\ {\mathrm H}(\omega )=\langle{\mathrm F}(\omega
)\rangle . \label{sb}
\end{eqnarray}
In our model
\begin{eqnarray}
{\mathrm K}(\omega )=\sum_{n}p_{n} |i_{n}(\omega )|^{2},\qquad
{\mathrm H}(\omega ) =\sum_{n}p_{n} i_{n}(\omega ).\label{model}
\end{eqnarray}
 The second term in Eq. (\ref{s}) results from the fluctuations of
the intervals among the arrival times of the pulses,
\begin{eqnarray}
{\mathrm S}_{c}(\omega )=\frac{8\pi^2}{\langle\tau\rangle}|{\mathrm H}(\omega )|^{2}
{\mathrm M}(\omega ),
\end{eqnarray}
where
\begin{eqnarray}
{\mathrm M}(\omega )=1+2\, \mathrm{Re} \,\frac{\phi (\omega )}{1-\phi
(\omega )}\, . \label{m}
\end{eqnarray}

There are various sources for the fluctuations  of the intervals
among the arrival times of the pulses, given by the distribution
function $w(\tau )$, see Eq. (\ref{phi}). One is the instability
in the SAW source, which is extremely small. Another, even more
important, is fluctuations originated from  the instability of the
gate voltages, which lead to fluctuations in the electron density
and consequently in the sound velocity, due to the renormalization
effect of the interaction between the SAW's and the 2DEG. In any
event, it is expected that the distribution $w(\tau)$ is very
narrow, such that the fluctuations,  $\delta\tau =\tau
-\langle\tau\rangle $, of those intervals are small, $\delta
\tau\ll \tau$. In other words,  $w(\tau)$ is  strongly peaked
around $\langle\tau\rangle=f^{-1}$. When the width of that peak is
ignored, that is, $w(\tau )=\delta \left(\tau
-\langle\tau\rangle\right )$, the function ${\mathrm M}(\omega )$
will vanish, unless  $\omega $ coincides with the harmonics of the
SAW frequency $\omega_{k}=2\pi kf$, i.e.
\begin{equation}
{\mathrm M}(\omega)=\frac{2\pi}{\langle\tau\rangle} \sum_{k\neq
0}\delta\left(\omega-\omega_{k}\right)\, . \label{M1}
\end{equation}
As a result, ${\mathrm S}_{c}(\omega )$ consists of a comb-like set of spikes.
When the dispersion of the interval distribution is taken into account,
these spikes are smeared. Their shapes can be found \cite{Rytov} by using
for the function $\phi $, Eq. (\ref{phi}), the approximation
\begin{eqnarray}
\phi (\omega )=e^{-i\omega\langle\tau\rangle}\left[
(1- \omega^{2}\langle (\delta\tau )^{2}\rangle/2\right]\, .
\end{eqnarray}
A straightforward calculation then shows that the width of the
$k$-th spike at its half-height is
\begin{equation}
(\delta \omega )_{k}/\omega_{k} = 2\pi k \,\langle
(\delta \tau)^2 \rangle /\langle\tau\rangle^2\, , \label{deltaok}
 \end{equation}
 and its height is given by
\begin{eqnarray}
{\mathrm
M}(\omega_{k})=
4/\omega_{k}^{2}\left\langle(\delta\tau)^2\right\rangle\, .
\end{eqnarray}
In-between the spikes ${\mathrm M}(\omega )$ is small; for example, at
the mid-point between successive spikes, where $\omega =2\pi
(k+1/2)f$, its value is
$2\pi^2(k+1/2)^2\langle(\delta\tau)^2\rangle/\langle\tau\rangle^{2}$.
Finally, at small frequencies one may write
$$
\phi (\omega )= 1-i\omega\langle\tau\rangle
-\omega^{2}(\left\langle (\delta \tau )^{2} \rangle
+\langle\tau\rangle^{2}\right)/2\, ,
$$ 
 to obtain
$
{\mathrm M(0)}=\langle (\delta\tau)^2\rangle/\langle\tau\rangle^{2}$.

Let us now discuss these results, in view of the experimental
findings. Near the first threshold of the acoustoelectric current,
it is plausible to assume that the potential wells capture either
one electron, or none at all. One then finds, using
$p_{0}+p_{1}=1$ in Eqs. (\ref{sb}) and (\ref{model}), that the
noise background is given by
\begin{eqnarray}
{\mathrm S}_{b1}(\omega )=8\pi^2 fp_{0}p_{1}|i_{1}(\omega )|^{2}.\label{sb0}
\end{eqnarray}
Similarly, near the second threshold, the wells capture either one
or two electrons, $p_{1}+p_{2}=1$, and one finds
\begin{equation}
{\mathrm S}_{b2}(\omega)=8\pi^2
fp_{1}p_{2}|i_{1}(\omega)-i_{2}(\omega)|^2.\label{sb1}
\end{equation}
In both situations the fluctuations are strongest at the
corresponding thresholds, where both relevant  probabilities are
close to 1/2. This is reminiscent of the shot noise in
point-contacts, which is proportional to ${\mathrm T}(1-{\mathrm
T})$ (where ${\mathrm T}$ is the transmission through the
point-contact) and has its maximum  at the conductance steps
\cite{shot}.

The spectral width, $\Delta\omega$, of the noise is determined by the
spectral width of the pulses $i_{n}(t)$. The duration of these pulses is
$a_{n}/s$, where $s$ is the SAW velocity and $a_{n}$ is the localization
length of $n$ electrons in the SAW well. This length depends on
microscopic details, but is expected to be some fraction, $\eta<1$, of the
SAW half-wavelength $s/2f$. Hence $\Delta\omega\simeq 2f/\eta$
(presumably, $\eta $ is small). At frequencies well below the width
$\Delta\omega $, $\omega\ll\Delta\omega$, one may then write $i_{n}(\omega
)\simeq ne/2\pi$, to find from Eqs. (\ref{sb0}) and (\ref{sb1})
\begin{eqnarray}
{\mathrm S}_{b1}(0)&=&2e^2 fp_{0}p_{1}\equiv 2e \langle
{\mathrm I}\rangle {\cal F}_{1}, \ \ {\cal F}_{1}=p_{0},\nonumber\\
 {\mathrm
S}_{b2}(0)&=&2e^{2} fp_{1}p_{2}\equiv 2e  \langle
I \rangle{\cal F}_{2},\ \ {\cal F}_{2}=\frac
{p_{1}p_{2}}{p_{1}+2p_{2}}.
\end{eqnarray}
Here, ${\cal F}_{i}$  are  the Fano factors, which indicate the
suppression of the noise compared to Poissonian noise. At the
thresholds the suppression is moderate, ${\cal F}_{1}=1/2$ and
${\cal F}_{2}=1/6$, while  far from them it is strong: ${\cal
F}_{1}=p_{0}\ll 1$ and ${\cal F}_{2}=p_{2}\ll 1$ below the
threshold and ${\cal F}_{2}=p_{1}/2\ll 1$ above.

As is mentioned above, the comb-like part of the spectrum,
${\mathrm S}_{c}$, consists of spikes located at $\omega_k= 2\pi
kf$, with an envelope given by $|{\mathrm H}(\omega_k)|^2$. The
width of this envelope is the same as that of the background,
$\Delta\omega$, which is much larger than the spike width
$\delta\omega_{k}$, and hence the spectrum contains about $2/\eta$
well-separated peaks. The amplitude of the $k$-th peak is
proportional to ${\mathrm M}(\omega_k)$.



The noise spectra thus provides one with information useful for the
understanding of the microscopic origin of the current quantization. In
particular, measuring the Fano factors, allows one to determine the
trapping probabilities $p_n$. Then for a trapping process which occurs
via tunneling, $p_{n}$ are expected to be temperature independent, while
temperature-dependent probabilities will indicate phonon-assisted trapping.

The spectral width of the current noise is related to the size of
the state, in which the electrons are trapped. Presumably, that
length will be longer for two electrons than for a single one,
$a_{2}>a_{1}$. This means that near the first threshold the noise
spectrum is broader, compared to the second one,
$(\Delta\omega)_{1}>(\Delta\omega)_{2}$.

In summary, we have proposed a phenomenological model  for the
fluctuations of the quantized acoustoelectric current, and derived
detailed predictions for the noise spectrum of the current. In particular,
we have found that measuring that noise gives additional information which
may turn out to be important in understanding the microscopic mechanism of
the current quantization.

\acknowledgements  This work has been done at the Centre of Advanced
Studies, Oslo, Norway, and  during the visit of YMG to the Weizmann
Institute of Science.
 Supports from the OEC
Project -- \textit{Access to Submicron Center for Research on
Semiconductor Materials, Devices and Structures} (HPRI-CT-1999-0026) and
the Israel Science Foundation are acknowledged. YMG thanks V. I.
Talyanskii for discussions related to the experiment.

\end{multicols}

\begin{references}

\bibitem{Wix1} A.~Wixforth, J.~P.~Kotthaus, and G.~Weimann,
\prl {\bf 56}, 2104 (1986);
 A.~Wixforth, J.~Scriba, M.~Wassermeir, J.~P.~Kotthaus, G.~Weimann,
 and W.~Schlapp, Phys. Rev. {\bf B} 40,  7874  (1989);
I. L. Drichko, A. M. Diakonov, I. Yu. Smirnov,  Y. M. Galperin,
and A.~I.~Toropov, \prb {\bf 62},
7470 (2000).
\bibitem{Wil}
 R.~L.~Willet, M.~A.~Paalanen, K.~W.~West, L.~N.~Pfeiffer, and
 D.~J.~Bishop, Phys. Rev. Lett. {\bf 65},  112  (1990);  R.~L.~Willet,
 R.~R.~Ruel,  M.~A.~Paalanen, K.~W.~West, and L.~N.~Pfeiffer,
Phys. Rev. {\bf B}  47, 7344 (1993).

\bibitem{Pepper} J.~M.~Shilton, D.~R.~Mace, V.~I.~Talyanskii,
M.~Pepper, M.~Y.~Simmons, A.~C.~Churchill, and D.~A.~Ritchie, Phys. Rev.
{\bf B} 51, 14770 (1995).
\bibitem{Sh1}

J. M. Shilton, D. R. Mace, V. I. Talyanskii, Yu. Galperin, M. Y. Simmons,
M. Pepper, and  D. A. Ritchie,
 J. Phys.: Condens. Matter {\bf 8}, L337 (1996).

\bibitem{Sh2}
J. M. Shilton, V. I. Talyanskii, M. Pepper, D. A. Ritchie, J. E. F. Frost,
C. J. Ford, C. G. Smith, and G. A. C. Jones,
 J. Phys.: Condens. Matter {\bf 8}, L531 (1996).

\bibitem{Ta97}
V. I. Talyanskii, J. M. Shilton,  M. Pepper, C. G. Smith, C. J.
Ford, E. H. Linfield, D. A. Ritchie, and G. A. C. Jones,
 Phys. Rev. {\bf B} 56, 15180  (1997).
\bibitem{AE1}  H. Totland and Y. M. Galperin, Phys. Rev. {\bf B} 54, 8814
 (1996); Physica scripta {\bf 69}, 302 (1997).
\bibitem{AE2} V. L. Gurevich, V. B. Pevzner, and G. J. Iafrate,
     Phys. Rev. Lett. {\bf  77}, 3881 (1996); V. L. Gurevich, V. I.
     Kozub, and
V. B. Pevzner, Phys. Rev. {\bf B} 58, 13088 (1998).

\bibitem{Thouless} D. J. Thouless, Phys. Rev. {\bf B}  27, 6083 (1983).
\bibitem{Aizin}
G. R. Aizin, G. Gumbs, and M. Pepper, Phys. Rev. {\bf B} 58, 10589 (1998).

\bibitem{Maksym}
P. A. Maksym, Phys. Rev. {\bf B} 61, 4727 (2000).

\bibitem{Flensberg} K.~Flensberg, Q.~Niu, and M.~Pustilnik, Phys. Rev. {\bf B}
60, R16 291 (1999).

\bibitem{Rytov} S. M. Rytov, Yu. Kravtsov, and V. I. Tatarsky, {\em
Principles of Statistical Radiophysics\/}, Springer, Berlin
(1989), part 2.

\bibitem{Kogan} Sh. Kogan, \emph{Electronic Noise and Fluctuations in
Solids}, Cambridge University Press (1996), p. 14.

\bibitem{end1}
Our result for ${\mathrm S}(\omega )$ is formally not valid  for $\omega=0$.
The zero frequency is, however, not relevant, since the lowest frequency
accessible in the experiment is larger than the inverse measuring
(averaging) time $T$. Only for such frequencies the random process can be
considered as a stationary one.

\bibitem{shot}
G. B. Lesovik, JETP Lett. {\bf 49}, 592 (1989).
\end{references}
\end{document}